\documentclass[lineno, 12 pt]{biometrika}

\usepackage{amsmath}
\usepackage{float}
 
\usepackage{lipsum}
\usepackage{bm}
\usepackage{natbib}
\usepackage[plain,noend]{algorithm2e}
\usepackage[utf8]{inputenc}
\usepackage{amssymb}
\usepackage[justification=centering]{caption}

\usepackage{graphicx}
\usepackage{mathtools}
\usepackage{booktabs}
\usepackage{amstext} 
\usepackage{epstopdf}
\usepackage{array} 
\nolinenumbers
\title{Connecting population-level AUC and latent scale-invariant $R^2$ via Semiparametric Gaussian Copula and rank correlations}

\author{D. Dey}
\affil{Johns Hopkins Bloomberg School of Public Health,\\ Baltimore, MD, USA \email{ddey1@jhu.edu}}

\author{V. Zipunnikov}
\affil{Johns Hopkins Bloomberg School of Public Health,\\ Baltimore, MD, USA
\email{vzipunn1@jhu.edu}}

\begin{document}

\maketitle

\begin{abstract}
Area Under the Curve (AUC) is arguably the most popular measure of classification accuracy. We use a semiparametric framework to introduce a latent scale-invariant $R^2$, a novel measure of variation explained for an observed binary outcome and an observed continuous predictor, and then directly link the latent $R^2$ to AUC. This enables a mutually consistent simultaneous use of AUC as a measure of classification accuracy and the latent $R^2$ as a scale-invariant measure of explained variation. Specifically, we employ Semiparametric Gaussian Copula (SGC) to model a joint dependence between observed binary outcome and observed continuous predictor via the correlation of latent standard normal random variables. Under SGC, we show how, both population-level AUC and latent scale-invariant $R^2$, defined as a squared latent correlation, can be estimated using any of the four rank statistics calculated on binary-continuous pairs: Wilcoxon rank-sum, Kendall's Tau, Spearman's Rho, and Quadrant rank correlations. We then focus on three implications and applications: i) we explicitly show that under SGC, the population-level AUC and the population-level latent $R^2$ are related via a monotone function that depends on the population-level prevalence rate, ii) we propose Quadrant rank correlation as a robust semiparametric version of AUC; iii) we demonstrate how, under complex-survey designs, Wilcoxon rank sum statistics and Spearman and Quadrant rank correlations provide asymptotically consistent estimators of the population-level AUC using only single-participant survey weights. We illustrate these applications using binary outcome of five-year mortality and continuous predictors including Albumin, Systolic Blood Pressure, and accelerometry-derived measures of total volume of physical activity collected in 2003-2006 National Health and Nutrition Examination Survey (NHANES) cohorts.
\end{abstract}


\section{Introduction} 

Classification offers a wide range of techniques from classical methods modelling underlying probabilities such as logistic regression and linear discriminant analysis to more recent methods including support vector machines, random forests, and neural networks \citep{kuhn2013applied, bishop2006pattern}. Various measures of classification accuracy have been proposed \citep{steyerberg2009clinical, steyerberg2010assessing, harrell2015regression}. Receiver Operating Characteristic curve (ROC) represents the accuracy of a classification model via a curve of trade-offs between false positive and true positive rates \citep{kuhn2013applied, saito2015precision}. The Area Under the ROC Curve (AUC) is often used to summarize the entire ROC curve and to compare classification models in terms of their discrimination strength. AUC can be equivalently represented as the probability that a randomly chosen case has a larger value of the continuous predictor than a randomly chosen control. From this conditional distribution perspective, AUC can be viewed as a fully nonparametric measure of concordance between an observed binary outcome and an observed continuous predictor. This definition has been linked to nonparametric Kendall's Tau rank correlation and Wilcoxon rank-sum statistics \citep{kendall1987kendall}. The limitations of AUC as a single summary of classification accuracy has been widely discussed in literature \citep{lobo2008auc, harrell2015regression, steyerberg2009clinical, saito2015precision}. 

In regression models with binary outcome such as logistic and probit regressions, an alternative fully parametric perspective is taken by using goodness-of-fit measures. A big class of these measures are likelihood-based \citep{tutz2011regression, demaris2002explained, schemper2003predictive}. Another class of these measures focuses on extending $R^2$, as a well-understood goodness-of-fit measure for linear models with continuous outcome. Specifically, two main alternative interpretations of $R^2$ as the proportion of variance explained and as a squared correlation have been adapted and extended to models with binary outcome by \cite{demaris2002explained, schemper2003predictive, yazici2007comparison}. 

We will use Semiparametric Gaussian Copula \citep{fan2017high} (SGC) to bridge these two perspectives and to connect AUC as a measure of classification accuracy and a novel latent scale-invariant $R^2$ as a measure of explained variation. Semiparametric Gaussian Copula (SGC) was used by \cite{fan2017high} to model a joint dependence between an observed binary outcome and an observed continuous predictor via the correlation of latent standard normal random variables. A major computational advantage of the approach in \cite{fan2017high} is the estimation procedure that does not require any likelihood maximization and estimates the latent correlation via a bridging procedure. The procedure links Kendall's Tau for binary-continuous pairs and the latent correlation via a known monotone "bridging" function that depends on the population-level prevalence rate of cases. The plug-in estimation is done using sample versions of Kendall's Tau and the prevalence rate. This bridging trick, also called inversion, is frequently used to estimate parameters of specific copula families via linking these parameters to Kendall's Tau or other rank correlations and inverting these links \citep{nelsen2007introduction, joe2014dependence}. Conceptually, the bridging argument in \cite{fan2017high} is similar to the to classical results bridging biserial correlation on binary-continuous pairs and triserial correlation on binary-binary pairs to Pearson correlation of underlying continuous variables that generated binary variables via dichotimization\citep{maccallum2002practice}. Because SGC uses Kendall's Tau to estimate the latent correlation, it becomes possible to connect AUC to the latent correlation capturing dependence between an observed binary outcome, generated via dichotomization of underlying latent continuous variable, and an observed continuous predictor. In addition to classical expression of AUC via Kendall's Tau and Wilcoxon rank sum statistics, we will show that AUC is a linear (up to an absolute value, here and throughout the paper) function of Spearman rank correlation \citep{sidak1999theory} for binary-continuous pairs. Then, under SGC, we will explicitly link these three rank statistics as well as Quadrant rank correlation \citep{sidak1999theory} to the latent $R^2$, defined as a square of the latent correlation, via corresponding monotone "bridging" functions that we derive in this paper. Importantly, being semiparametric our approach results in the latent $R^2$ that is scale-invariant. 

After building the framework, we will focus on three applications. The first application will illustrate a significant added value of the joint and mutually consistent interpretation of AUC as a measure of classification accuracy and the latent $R^2$ as a measure of the proportion of variance explained between an observed binary outcome and an observed continuous predictor. In the second application, we will propose to use Quadrant rank-correlation as a robust semiparametric version of AUC. As shown in \cite{croux2010influence}, Quadrant rank correlation is more robust compared to Kendall's Tau and Spearman rank correlations in continuous-continuous case. In extensive simulation studies, we show that a similar robust performance is observed when sample is contaminated with outliers in binary-continuous case. In the third application, we will demonstrate how our framework addresses a problem of calculating AUC under complex survey designs. Specifically, we will show how Wilcoxon's rank-sum statistics as well as Spearman and Quadrant rank correlations can be used with single participant survey weights to construct asymptotically unbiased estimators of the population-level AUC.

The rest of the paper is organized as follows. In Section 2, we discuss the four rank-statistics and their relationship with AUC. In Section 3, we introduce Semiparametric Gaussian Copula and derive bridging functions that connect the four rank statistics to the latent $R^2$. In Section 4, we discuss the main applications of the framework. In Section 5, we provide extensive simulation results. In Section 6, we demonstrate the proposed framework on NHANES 2003-2006 cohorts. Discussion concludes with a summary, limitations, and future work.

\section{AUC and Rank Statistics} \label{sec:auc-rank}

In this section, we establish the links between AUC and the rank statistics. Let us first introduce notations. Throughout the paper, we will consider binary-continuous pairs of random variables $(Y,X)$, where $Y$ is an observed binary outcome and $X$ is an observed continuous predictor. We will refer to $M_Y, M_X$ as the population medians of $Y$ and $X$, respectively. We denote by $F_Y(), F_X()$ the cumulative distribution functions of random variables $Y$ and $X$, respectively. We will refer to $Y=1$ as a case and define the population-level prevalence rate of cases as $p = P(Y=1)$. Finally, we denote $X_1 =(X|Y=1)$ and $X_0 = (X|Y=0)$ to be random variables following the conditional distribution of the continuous predictor for cases, $Y=1$, and controls, $Y=0$, respectively. Using these notations, population-level AUC, denoted by $A$, can be defined as $A = \max\{P(X_1 > X_0), P(X_1 < X_0)\}$. It is easy to see that $P(X_1 > X_0) = 1- P(X_1 < X_0)$, so $A \geq 0.5$. 

We consider three rank correlations including Kendall's Tau, Spearman's Rho and  Quadrant, also known as Blomqvist's Beta, as well as Wilcoxon's rank-sum statistic, used to nonparametrically test the equality of two distributions. We lay out the population-level definitions of these rank statistics in the case of binary-continuous pairs: 1) \textbf{Kendall's Tau}: $r_K=E((Y_i - Y_i') sgn(X_i - X_i'))$; 2) \textbf{Wilcoxon's rank-sum statistic}: $W=P(X \leq X_1) - P(X \leq X_0)$; 3) \textbf{Spearman correlation}: $r_S=12E[F_Y(Y)F_X(X)] - 3$; 4)\textbf{Quadrant correlation}: $r_Q=E[sgn((Y-M_Y)(X-M_X))]$, where $(Y_i,X_i)$ and $(Y_i^{'},X_i^{'})$ are two independent copies following the same bivariate joint distribution and $sgn(x)=I\{x\neq 0\}(2I\{x>0\}-1)$ denotes the sign function. We next establish the relationship between these rank statistics and AUC. All derivations for these results are presented in Appendix \ref{appendix: auc-rank}. 

Kendall's Tau captures concordance within a pair of bivariate observations and relates to AUC as $A = 0.5 + \left|r_K/(4p(1-p))\right|$. 

Wilcoxon rank-sum statistic is a linear rank statistics. The use of $W$ for calculating AUC under complex survey designs has been discussed by Professor Thomas Lumley in his blog \emph{Bias and Inefficient} (https://notstatschat.rbind.io/2017/12/26/statistics-on-pairs/). Additionally, \cite{lumley2013two} demonstrated how survey-weighted rank test can be constructed using Wilcoxon rank-sum statistic to compare two distributions in a complex survey design. We followed his approach by fixing a minor error and obtaining population-level relationship $A= 0.5+|W|$. 

Finally, we show that Spearman's rank correlation and Wilcoxon rank-sum are linearly related in binary-continuous case. This is primarily due to the fact that the cumulative distribution function of a binary random variable takes exactly two values. Specifically, we show that Spearman's rank correlation and AUC are linearly related as follows $A = 0.5 +\left|(r_S - (6p^2 - 6p+3))/(12p^2(1-p))\right|$. 

Even though we can relate AUC linearly to the three rank statistics above, it is not possible to express Quadrant correlation linearly in terms of AUC. The calculation of Quadrant correlation requires counting the numbers of pairs in the first, second, third, and fourth quadrant of the $(X,Y)$ two-dimensional plane. As it involves only the sign of the distance of ranks from the median and not the ranks themselves, Quadrant is more robust but less efficient than Kendall's Tau and Spearman's Rho correlations in continuous-continuous case \citep{croux2010influence}. As we will show in the next section, under semiparametric Gaussian copula assumption, we can non-linearly relate Quadrant rank correlation to AUC and the other three rank statistics. 

\section{Semiparametric Gaussian Copula}

\subsection{Introduction to the copula}

The assumption of multivariate Gaussinity is arguably the most popular in multivariate statistical analysis. However, in many applications, this assumption is not realistic. To address this, \cite{liu2012high} proposed a Semi-parametric Gaussian Copula (SGC) model. Below, we provide definitions from \cite{liu2012high}.

\begin{definition}
We say that a pair of continuous random variables $(Y,X)$ follows a \textbf{non-paranormal} distribution, if there exist monotone functions $f_Y(), f_X()$ such that $(U,V)=(f_Y(Y),f_X(X)) \sim N_2(0,0,1,1,r)$. 
\end{definition}
For binary-continuous pairs, \cite{fan2017high} defined latent non-paranormal distribution as follows. 
\begin{definition}
Suppose we have binary variable $Y$ and continuous variable $X$. Then, if there exists a latent variable $Z$ and monotone functions $f_Z(),f_X()$ such that $(Y,X) = (I\{f_Z(Z) > \Delta\}, X)$ and $(U,V)=(f_Z(Z),f_X(X)) \sim N_2(0,0,1,1,r)$, then we say that the binary-continuous pair $(Y,X)$ follows \textbf{latent non-paranormal distribution}. 
\end{definition} 

The approach in \cite{fan2017high} estimates the latent correlation via mapping Kendall's tau using a one-to-one function $G_K()$, called "bridging" function, so that $r_{K}= G_K(r)=4\Phi_2(\Delta,0,\frac{r}{\sqrt{2}}) - 2 \Phi(\Delta)$, where $\Phi_2(a,b,r)$ denotes the cumulative distribution function of a standard bivariate normal distribution with correlation $r$. Bridging function $G_K(r)$ is odd, so $G_K(-r)=-G_K(r)$. Figure \ref{fig:rkvsr} in Appendix \ref{appendix: fig} shows $G_K(r)$, $G_K^{'}(r)$, and $G_K^{''}(r)$ for different values of $p = 1 -\Phi(\Delta)$. Based on the sign of $G_K^{''}(r)$, we can see that $G_k(r)$ is convex for higher values of $p$ and concave for lower values of $p$, and neither in between.

Thus, it becomes possible to connect AUC, via Kendall's Tau and SGC assumption, to the latent correlation that captures the dependence at data generating level between an observed binary outcome, conceptualized as a dichotomized continuous variable, and an observed continuous predictor. 

\subsection{Bridging the latent correlation and rank statistics} \label{sec:sgc-bridge}
Next, we will show how Spearman and Quadrant rank correlations can also be used to estimate the latent correlation of SGC.

\begin{lemma}\label{lemma: sgc-bridge}
Under SGC, Spearman rank correlation, $r_S$, and Quadrant rank correlation, $r_Q$, can be mapped to the latent correlation $r$ as follows:

\begin{equation}
\begin{split}
& r_S = G_S(r)=12 [\Phi_2(0,-\Delta,\frac{r}{\sqrt{2}}) + p \Phi_2(0,\Delta,-\frac{r}{\sqrt{2}})] - 3\\
& r_Q = G_Q(r)= [\Phi_2(-\Delta,0,r) - \Phi_2(-\Delta,0,-r)] \mathbb{I}(M_Y=0) + \\
& \qquad [\Phi_2(\Delta,0,r) - \Phi_2(\Delta,0,-r)] \mathbb{I} (M_Y=1).\\
 \end{split}
\end{equation}

\end{lemma}

The detailed proof is provided in Appendix\ref{appendix:bridge}. 

We will refer to $G_K(),G_S(),G_Q()$ as bridging functions and the subscripts $K, S, Q$ will specify a specific rank correlation. \cite{fan2017high} showed that $G_K(r)$ is a strictly increasing function of $r$ in $(-1,1)$. We establish similar results for $G_S(r)$ and $G_Q(r)$ in below.  

\begin{lemma} \label{lemma: bridge-inv}
The bridging functions $G_S(r)$ and $G_K(r)$ are strictly increasing functions of $r$ in $(-1,1)$ and hence, the inverse functions exist.
\end{lemma}

The proof is provided in the Appendix\ref{appendix:bridge}. 

Thus, we can ensure that latent correlation estimators obtained by inverting bridging functions are well defined. It is important to remember a few properties of the bridging functions. First, the bridging function is constructed assuming independence at the population level. Second, we only use bridging functions to bridge between population-level statistics. Third, the bridging functions do not depend on the sampling scheme which might bring dependence between subjects through sampling mechanism.

\section{Applications}\label{sec: applications}
In this section, we consider main applications of the proposed framework.

Using the results from previous section, we now can connect the population-level AUC and rank statistics as follows.
\begin{equation}
\begin{split}
& {A}_K =  \frac{1}{2}+\left|\frac{{r}_K}{4p(1-p)}\right|,\\
& {A}_W =  \frac{1}{2}+|{W}|\\
& {A}_S  = \frac{1}{2} + \left|\frac{G_K(G_S^{-1}(r_S))}{4p(1-p)}\right| = \frac{1}{2}+ \left|\frac{r_S - (6p^2 - 6p+3)}{12p^2(1-p)}\right|, \\
& {A}_Q =  \frac{1}{2}+ \left|\frac{G_K(G_Q^{-1}({r}_Q))}{4p(1-p)}\right|.
 \end{split}
   \label{eqn: auc-rank}
\end{equation}
At the population-level, all four ways are equivalent, i.e. $A = A_K = A_W = A_S = A_Q$. Of course, in sample we likely end up with different estimates of AUC. It is also important to note that the dependence of AUC and the latent correlation $r$ involves the prevalence rate $p$.



\subsection{Latent $R^2$ for univariate continuous predictor}

Our approach provides a framework to introduce a novel goodness-of-fit statistics, the latent $R^2$, that we denote as $R_l^2$ and define as a square of the latent correlation. Various goodness-of-fit measures have been previously proposed for models with binary outcome. Many of them focused on extending $R^2$, as a popular and intuitive measure in linear models with continuous outcome. Two main alternative interpretations of $R^2$ i) as the proportion of variance explained or ii) as squared correlation have been pursued and generalized to models with binary outcome \cite{demaris2002explained, schemper2003predictive, yazici2007comparison}. Among the limitations of those proposals is that the range of the values can go outside of the usual $(0,1)$ interval as well as the lack of invariance to the scale of the continuous predictors \cite{demaris2002explained, schemper2003predictive, yazici2007comparison}. 

Under SGC, we estimate the latent correlation using three different estimators corresponding to three rank correlations:  
\begin{equation}
\begin{split}
 R^2_{lK} = (G_K^{-1}(r_K))^2,\\
 R^2_{lS} = (G_S^{-1}(r_S))^2,\\
 R^2_{lQ} = (G_Q^{-1}(r_Q))^2.
 \end{split}
   \label{r-sq}
\end{equation}
Again, at the population level, $R^2_l = R^2_{lK}= R^2_{lS} = R^2_{lQ}$.

Thus, $R^2_l$ quantifies the proportion of variance explained by the continuous predictor in the latent normalized space and gives us back a familiar intuition available for linear models with a continuous outcome. 

\begin{figure}[htb!]
\centering
\includegraphics[width=160mm, height=70mm]{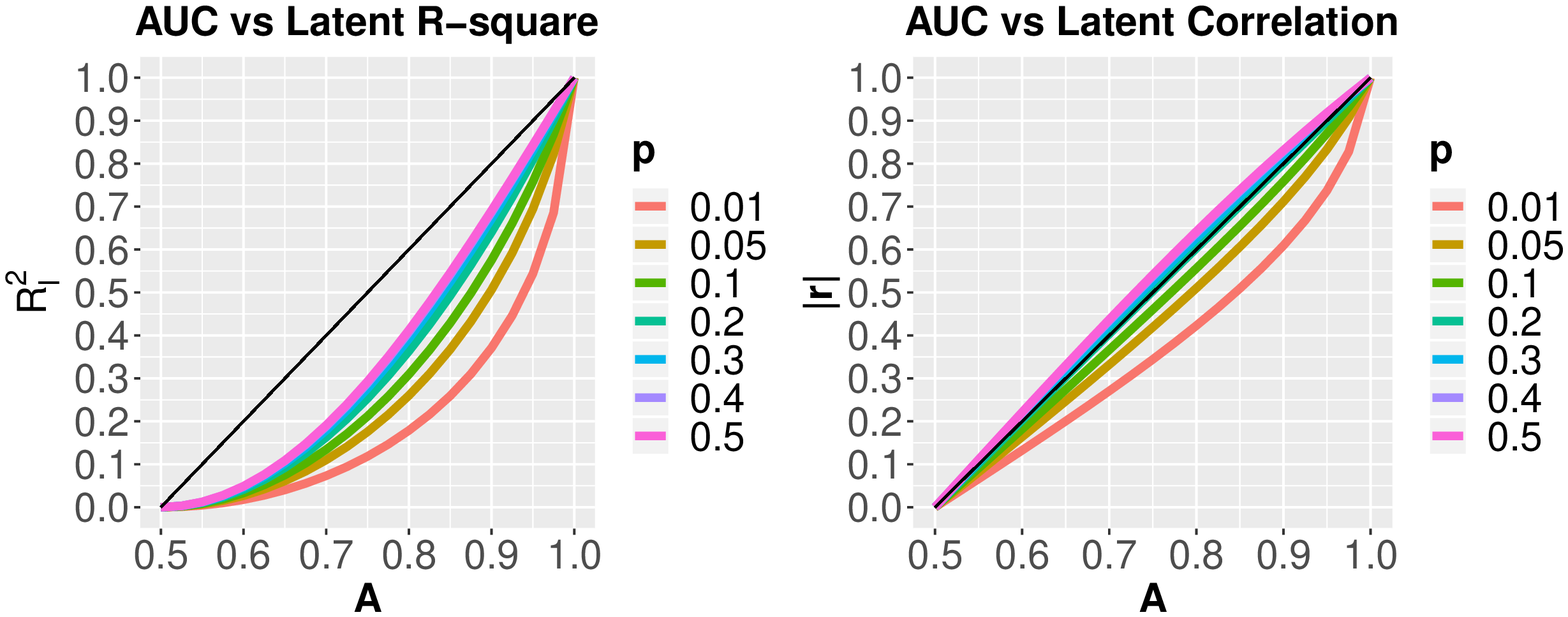}
\caption{The relationships between AUC and the latent $R^2_l$ (left panel) and AUC and the absolute value of latent correlation (right panel) and its dependence on $p$.}
\label{fig:aucvsr2}
\end{figure}

Figure \ref{fig:aucvsr2} shows the relationship between AUC and the latent $R^2_l$ in the left panel and the relationship between AUC and the absolute value of the latent correlation, $|r|$, in the right panel. As a reference, we include a linear line $|r|=2A-1$. We show these relationships for different values of the prevalence rate, $p$. The right panel shows that AUC and $|r|$ are almost identical for $p = 0.5$, but with $p$ getting smaller, the same value of AUC corresponds to increasingly smaller values of $|r|$.  For example, AUC of $0.8$ corresponds to $|r|$ of $0.6$, if $p = 0.5$, and $|r|$ of $0.4$, if $p = 0.01$. The latent correlation tends to lie below the linear reference line for most of values of $p$. The same observations are true for the relationships between AUC and $R^2_l$, but with a much larger curvature due to the squared nature of $R^2_l$'s scale compared to the linear scale of $|r|$. 

\subsection{Latent $R^2$ for multivariate continuous predictor}
The definition of $R^2_l$ can be readily extended to a case of a linear combination $X'\beta$ for a fixed pre-defined multivariate parameter $\beta$. Indeed, a linear combination $X'\beta$  can be treated as a scalar continuous predictor and, if the SGC assumptions for $X'\beta$ are valid, the same argument connecting AUC and $R^2_l$ can be applied to a binary outcome $Y$ and a scalar continuous predictor $X'\beta$. 

An alternative two-step procedure to define $R^2_l$ for a multivariate continuous predictor can be done as follows. First, estimate the joint latent correlation matrix, $\Sigma$, of binary outcome $Y$ and multivariate continuous predictor $X$. Second,
define $R^2_l$ as $R^2_l = \Sigma_{YX}\Sigma_{XX}^{-1} \Sigma_{XY}$, where $\Sigma_{YX}$ denotes the latent correlation between the outcome and predictor and $\Sigma_{XX}$ denotes the latent correlation matrix for the predictor. This approach generalizes $R^2$ for linear models with continuous outcome and continuous predictors. However, whether there is a specific relationship between AUC and $R^2_l$ defined in this way remains to be investigated and is beyond the scope of this paper.

\subsection{Complex survey designs} \label{sec: survey}
Many social and public health studies are conducted using data from surveys following complex designs. To properly estimate the population-level AUC in complex surveys, one would need to know the pairwise sampling weights, or pairwise probabilities of selection, for study participants. However, pairwise survey weights are often not available. To tackle this, researchers have used various approximations such as the product of individual participant weights \citep{korn2011analysis} or have applied non-parametric approaches \citep{yao2015estimation}. In this section, we show how rank statistics and single participant survey weights can be used to construct asymptotically unbiased estimators of the population-level AUC.

First, we introduce the definition of the survey-weighted AUC. 

\begin{definition}
Suppose, $X_{11}, X_{12},...,X_{1m_1}$ and $X_{01}, X_{02},...,X_{0m_0}$ are i.i.d. samples from the distributions of $X_1$ and $X_0$, respectively. Let us assume that the total sample size is $n = m_0 + m_1$. Then, the survey-weighted AUC can be calculated as
\begin{equation}\hat{A}_{wt}= \frac{1}{{\sum_{i=1}^{m_0}\sum_{j=1}^{m_1}\frac{1}{w(i,j)}}}   \sum_{i=1}^{m_1}\sum_{j=1}^{m_0} \frac{1}{w(i,j)}h(X_{1i},X_{0j})=\hat{E}_wh(X_1,X_0),
\label{wtauc}
\end{equation} 
where $h(x,y) = \mathbb{I}(x>y) + 0.5\mathbb{I}(x=y)$, $w(i)$ and $w(i,j)$ are single and pairwise participant weights, respectively. \end{definition}

The key idea of the proposal outlined below is to use single participant survey weights to estimate population-level rank statistics and then connect them to the population-level AUC using bridging functions. We can define the survey-weighted estimators (Horvitz-Thompson estimators) of Kendall's Tau, Wilcoxon rank-sum statistic, Spearman and Quadrant rank correlations as follows.

\begin{equation}
\begin{split}
& \hat{r}_K  = \frac{1}{{\sum_{i<j}\frac{1}{\hat{w}(i,j)}}} \sum_{i < j} \frac{1}{\hat{w}(i,j)} [(Y_i - Y_j) sgn(X_i - X_j)]\\
&\hat{W} = \frac{1}{\sum_{i: Y_i=1}\frac{1}{w(i)}}\sum_{i: Y_i=1} {\frac{1}{w(i)}\hat{F}_X(X_i)} - 
\frac{1}{\sum_{i: Y_i=0}\frac{1}{w(i)}}\sum_{i: Y_i=0} {\frac{1}{w(i)}\hat{F}_X(X_i)}\\
& \hat{r}_S  = 12  \frac{1}{{\sum_{i=1}^{n}\frac{1}{w(i)}}} \sum_{i=1}^{n} \frac{1}{w(i)} [\hat{F}_Y(Y_i)\hat{F}_X(X_i)] - 3\\
& \hat{r}_Q  = \frac{1}{{\sum_{i=1}^{n}\frac{1}{w(i)}}}\sum_{i=1}^{n} \frac{1}{w(i)} sgn((Y_i-\hat{M}_Y)(X_i-\hat{M}_X)).
\end{split}
\label{eqn: k-w-s-q-survey}
\end{equation} 

The estimates of population-level medians, $M_Y$ and $M_X$, and the population-level distribution functions, $F_Y$ and $F_X$, are obtained using Horvitz-Thomposon estimators.

\cite{lumley2013two} established the asymptotic properties of statistics of the form 

$$\hat{T}= \frac{1}{\sum_{i: Y_i=1}\frac{1}{w(i)}}\sum_{i: Y_i=1} {\frac{1}{w(i)}g(\hat{F}_X(X_i)}) - 
\frac{1}{\sum_{i: Y_i=0}\frac{1}{w(i)}}\sum_{i: Y_i=0} {\frac{1}{w(i)}g(\hat{F}_X(X_i)})$$

under complex survey designs. The function $g()$ can follow any of the assumptions stated in Appendix \ref{appendix: asymp-qs}. Theorem $1$ in \cite{lumley2013two} has been used to show that $\sqrt{n}(\hat{W}-W)$ are asymptotically normal with mean zero. We adopt a similar approach to prove the asymptotical normality for $r_S$ and $r_Q$.

\begin{theorem}\label{thm: asymp-qs}
Under assumptions $A1$ to $A4$ in Appendix \ref{appendix: asymp-qs}, $\sqrt{n}(\hat{r}_S - r_S)$ and $\sqrt{n}(\hat{r}_Q - r_Q)$ are asymptotically normal with mean zero. Hence, $(\hat{r}_S - r_S)$ and $(\hat{r}_Q - r_Q)$ converge to zero in probability. 
\end{theorem}

The proof is provided in Appendix \ref{appendix: asymp-qs}. 

Kendall's Tau requires pairwise survey weights $\hat{w}(i,j)$. We consider three different estimates of Kendall's Tau: (1) \emph{unweighted}, $\hat{r}_{Kuw}$, with $\hat{w}(i,j)=1 $; (2) \emph{true weighted}, $\hat{r}_{Ktw}$, with $\hat{w}(i,j)= w(i,j)$; and (3) \emph{product weighted}, $\hat{r}_{Kpw}$, with $\hat{w}(i,j)= w(i)w(j) $. Note that in most of practical settings, we do not know or have an access to true pairwise weights and can only calculate $\hat{r}_{Ktw}$ in simulations. 

We can estimate the population-level AUC using Equations (\ref{eqn: auc-rank}) and define corresponding estimates as $\hat{A}_{Kuw}$, $\hat{A}_{Ktw}$, $\hat{A}_{Kpw}$, $\hat{A}_{W}$, $\hat{A}_{S}$, and $\hat{A}_{Q}$.
 
Using Theorem \ref{thm: asymp-qs} and applying the delta method to Equations \ref{eqn: auc-rank} leads to the the following result. 

\begin{corollary}
Under the assumptions of Theorem \ref{thm: asymp-qs}, $\sqrt{n}(\hat{A}_S - A_S)$ and $\sqrt{n}(\hat{A}_Q - r_Q)$ are asymptotically normal with mean zero. Hence, $(\hat{A}_S - A_S)$ and $(\hat{A}_Q - A_Q)$ converge to zero in probability. 
\end{corollary} 

Note that we can apply delta method because of the differentiability of the bridging functions as shown in the proof of Lemma \ref{lemma: bridge-inv} in Appendix.

Thus, using single participant survey weights, we can define asymptotically unbiased estimators of the population-level AUC. Finally, we can invert bridging functions to get three different estimates of the latent $R_l^2$ as $\hat{R}_{lK}^2 $, $\hat{R}_{lS}^2$, and $\hat{R}_{lQ}^2$, respectively. 

\section{Simulations}


In this section, we perform extensive simulation studies to explore the behaviour of the proposed estimators of $AUC$. Under two-stage stratified cluster sampling, we consider several different combinations of strata informativeness and outlyingness in binary-continuous pairs.

Following \cite{yao2015estimation}, we set up a two-stage stratified cluster sampling as follows. Suppose, we have a finite population, $N$ subjects, with $H$ strata and each strata has $K$ Primary Sampling Units (PSUs). The number of subjects in the $g$th PSU in the $h$-th strata are $N_{hg}$. Suppose, we pick $n_{hg}$ subjects from each PSU $hg$ and the total number of subjects selected from $h$th strata is $n_h$ with the total number of subjects in the sample to be $n$. We sample subjects as follows: first, randomly select $u$ out of $U$ PSUs within each stratum and then randomly select a fixed number of subjects ($n_h/u$) from each of the PSU in $h$-th strata, for $h=1,\cdots, H$. Based on this sampling scheme, we can easily calculate both the individual selection probability and pairwise selection probabilities as described in Section 2.3 of \cite{yao2015estimation}. 

\begin{figure}[htb!]
\centering
\includegraphics[width=160mm, height=100mm]{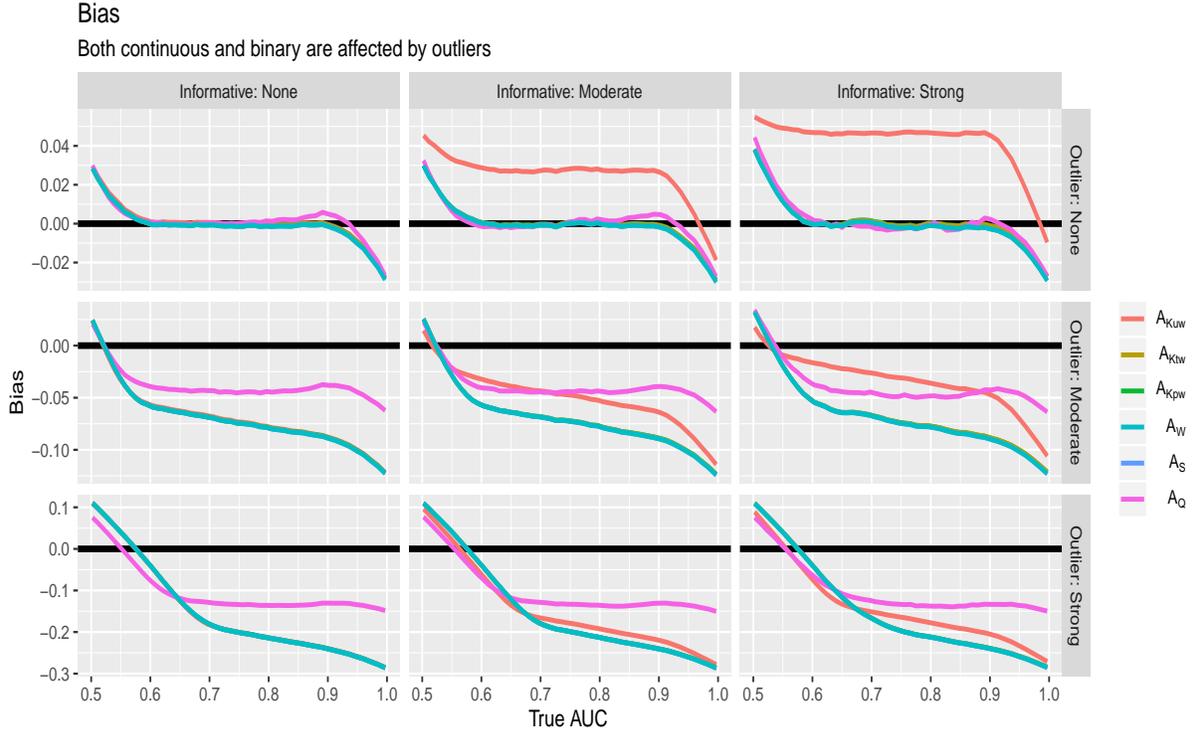}
\caption{Simulation results: Bias of the AUC estimators under different scenario}
\label{fig: bias-sim1}
\end{figure}


 First, we define \textbf{strata informativeness} as follows. If $n_h$ depends on the AUC in $h$-th strata, we call this informative sampling. If the sampling is informative, we oversample strata with higher AUCs. We rank strata according to AUC for that specific strata. Depending on the level of oversampling, we define three types of informativeness: 0-None ($n_h = n/H$ for all $h$) , 1-Moderate (top three: $n_h=n*0.18$, middle four: $n_h=n*0.07$, bottom three: $n_h=n*0.06$), 2-Strong (top three: $n_h=n*0.22$, middle four: $n_h=n*0.07$, bottom three: $n_h=n*0.02$). We define {\bf outlyingness} as follows. Following \cite{croux2010influence}, we introduce ``outlying" observations at latent level by substituting $N_2(0,0,1,1,r)$ with $N_2(4,-4,0.01,0.01,0)$ or $N_2(-4,4,0.01,0.01,0)$ randomly selected with probability $\frac{1}{2}$. We consider three levels of outlyingness: 0-None ($0\%$ of outliers), 1-Moderate ($5\%$ of outliers), 2-Strong ($15\%$ of outliers). 

We create $9$ different scenarios corresponding to the $3\times3$ combinations of strata informativeness and outlyingness. We take $H=10, U=10, u=2, n=600, N=60000$. For each scenario, we simulate data and calculate estimates as follows. 
\vspace{0.3in}
\setcounter{step}{0}

(a) \textbf{Schematics} {
\begin{step}
For a fixed latent correlation, $r$, we calculate true AUC, $A$, using bridging.
\end{step}
\begin{step}
 Take strata-specific AUCs, $A_h$, equally spaced within $(A-0.1, A+0.1)$. For example, if AUC $A$ is $0.8$ and there are 10 stratas, the strata specific AUCs are $0.7, 0.722, 0.744, 0.766, 0.788, 0.811, 0.833, 0.855, 0.877, 0.9$.
 \end{step}
 \begin{step}
 Depending on informativeness of the sampling scenario, calculate $n_h$ based on $A_h$.
 \end{step}
  \begin{step} 
    Calculate individual, $w(i)$, and pairwise, $w(i,j)$, survey weights for this particular scheme. 
\end{step}
}
\setcounter{step}{0}

(b) \textbf{Within Strata}
\begin{step} 
     For $A_h$, use bridging to calculate strata-specific latent correlation, $r_h$.
     \end{step}
     \begin{step} 
  Generate $(U_i,V_i),i=1,\cdots,N_h \sim N_2(0,0,1,1,r_h)$.
  \end{step}

\setcounter{step}{0}
(c) \textbf{Sampling and outliers}
\begin{step} 
Get a sample of size $n$ using sampling scheme described in the beginning of the section.
\end{step}
\begin{step}
Depending on the level of ``outlyingness", randomly select $0\%$, $5\%$ or $15\%$ of ``outliers'' across all stratas by changing corresponding $(U_i,V_i)$ from $N_2(0,0,1,1,r_h)$ to a random draw from $N_2(4,-4,0.01,0.01,0)$ or $N_2(-4,4,0.01,0.01,0)$ with probability $\frac{1}{2}$. 
\end{step}

\begin{step} 
Transform $Y_i=I(U_i > \Delta)$ and keep $X_i=V_i$ for the final finite sample. 
\end{step}

\setcounter{step}{0}

(d) \textbf{Estimation} 
     \begin{step} 
     Estimate $\Delta$ as $\hat{\Delta}=\Phi^{-1}(1-\frac{\sum_{i=1}^{n}\frac{Y_i}{w_i}}{\sum_{i=1}^{n}\frac 1 w_i})$
     \end{step}
     \begin{step} 
Estimate $\hat{r}_K$, using three different weighting schemes, $\hat{r}_S$ and $\hat{r}_Q$, using single participant weights, by inverting corresponding bridging functions described in Section \ref{sec:sgc-bridge}. 
 \end{step}
 \begin{step} 
Estimate $A$ using Equations \eqref{eqn: auc-rank} to obtain $\hat{A}_{Kuw}$, $\hat{A}_{Ktw}$, $\hat{A}_{Kpw}$, $\hat{A}_{W}$, $\hat{A}_{S}$, and $\hat{A}_{Q}$.
 \end{step}

We vary true latent correlation $r$ between $(0.005,0.995)$ taking $50$ equally spaced points and generate data from a specific scenario $100$ times. The results are used to report bias and mean squared errors (MSEs). 

Figure \ref{fig: bias-sim1} that shows bias across the nine scenarios. Note that each row has row-specific range for vertical axis, with the lowest range in the top row ($0\%$ of outliers) and the highest range in the bottom row ($15\%$ of outliers). The key findings are as follows. With increasing informativeness in sampling (from the most left to the most right column), bias of the unweighted estimator is relatively higher than that of the weighted versions. With increasing outlyingness (from the top to the bottom row), bias of $A_Q$ is significantly lower than weighted and unweighted AUC obtained using the other rank correlations. Even in the most challenging scenario, shown in the bottom most right corner, $A_Q$ exhibits the maximum absolute bias of only $0.12$ while other estimates have absolute bias as large as $0.3$. This provides a clear illustration that $A_Q$ can be seen and used as a robust semiparametric version of AUC. In the case of higher sampling informativeness and moderate presence of outliers, we observe that the unweighted AUC performs better than the weighted AUC. We argue that the intuition behind this is that in these kind of scenarios, the outlying observations might have higher weights and thus, unweighted estimates might be less biased than the weighted ones. 

The similar observations are true for MSEs of the estimators, shown in Figure \ref{fig:mse} in Appendix \ref{appendix: fig}.

\section{Classification of 5-year mortality in NHANES 2003-2006}

The National Health and Nutrition Examination Survey (NHANES) is a biennial stratified multi-stage sample survey of noninstitutionalized US population conducted by Center of Disease Control and Prevention, USA (https://www.cdc.gov/nchs/nhanes/index.htm). Using complex survey weighting techniques, the results obtained from NHANES can be considered as nationally representative. To illustrate our approach, we use accelerometry and laboratory data as well as linked mortality data for NHANES 2003-2006. We define a 5-year all cause mortality to be the observed binary outcome and use continuous predictors including age, albumin, systolic blood pressure and summaries of accelerometry-estimated physical activity which was a part of NHANES 2003-2006 protocol \citep{leroux2019organizing}.

Age (in years) is derived from the NHANES variable \emph{RIDAGEYR}, defined as the age of a participant at the time of household screening. Albumin (in ug/mL) is the variable \emph{URXUMA}. Systolic blood pressure (in mmHg) has been calculated as the average of multiple readings of the systolic blood pressure (up to four sequential readings per participant) which we denote here as $BPXSY$. Following \cite{leroux2019organizing}, we also include total activity count (TAC), time in minutes spent in moderate-to-vigorous activity (MVPA) and active-to-sedentary transition probability (ASTP). These accelerometry-derived measures have been shown to be significantly predictive of a 5-year follow-up mortality with classification performance comparable to and sometimes exceeding that of Age \citep{leroux2019organizing, smirnova2019predictive}. We excluded participants who (1) have missing mortality information or alive with follow-up less than 5 years, (2) are younger than 50 or aged 85 and older, (3) have missing any of predictor variables of interest, (4) have died due to accident, and (5) had fewer than 3 days of data with at least 10 hours of estimated accelerometry wear time \citep{leroux2019organizing}. Our final analytical sample consisted of $3069$ subjects with $321$ deaths within $5$ years. This gives an unweighted estimate of prevalence rate as $p=0.1$. Note that we use term ``prevalence rate" while referring to $p$ loosely here, just to stay consistent with the rest of the paper. We use \emph{rnhanesdata} package to process and recalculate the survey weights for our chosen subset of participants. Finally, the binary outcome is defined as a binarized 5-year follow-up mortality.

We use methods described in Section \ref{sec: survey} and Equation \ref{eqn: k-w-s-q-survey} to calculate our proposed estimates and to calculate the standard error and confidence intervals of those estimates. Specifically, (i) we follow the approach by \cite{yeo1999bootstrap} and use bootstrap to get standard error and confidence intervals, (ii) we use \emph{svyrepdesign} function in \textbf{R} package \emph{Survey} to create $100$ bootstrap replicates of the sample design using survey weights, (iii) then we calculate AUC and $R^2_l$ estimates from each of these replicates, and report bootstrap standard error and $95\%$ confidence intervals in Tables \ref{tab: auc-nhanes} and \ref{tab: rsq-nhanes}.  

The results are consistent with those in  \cite{leroux2019organizing}. For unweighted AUC, $A_{Kuw}$, if we rank AUCs from highest to lowest, the predictors are ordered as TAC, Age, MVPA, ASTP, Albumin, Systolic BP. For  weighted versions of AUC, $A_{Kpw},A_{W}$, and $A_{S}$, MVPA becomes more discriminative than Age. However, Age has been used for NHANES survey weights, so an influence on the results is expected. Interestingly, when using $A_Q$, MVPA becomes the predictor with highest AUC and TAC becomes the second highest. This is likely due many elderly participants having zero minutes of moderate-to-vigorous physical activity, where as, TAC provides a better discrimination between participants with MVPA zeroes \citep{varma2018total}. Hence, because Quadrant rank correlation only counts the quadrant concordance, MVPA gets this slight preference by Quadrant.  

The results for $R^2_l$ show that TAC explains about $33\%$, $29\%$, $29\%$, and $23\%$ variation, when we use the unweighted and weighted Kendall's Tau, weighted Spearman's and Quadrant rank correlations, respectively. The lowest $R^2_l$ is for Systolic BP, which explains less than one percent of the variation.  

\begin{table}[htb!]
\centering
\resizebox{\textwidth}{!}{\begin{tabular}{|r|l|l|l|l|l|l|l|l|l|l|l|}
  \hline
 & Variables & $A_{Kuw}$ & Rank & $A_{Kpw}$ & Rank & $A_W$ & Rank & $A_S$ & Rank & $A_Q$ & Rank \\ 
  \hline
1 & TAC & 0.75 (0.75, 0.75) & 1 & 0.8 (0.75, 0.83) & 1 & 0.8 (0.75, 0.83) & 1 & 0.8 (0.75, 0.83) & 1 & 0.77 (0.73, 0.8) & 2 \\ 
  2 & MVPA & 0.73 (0.73, 0.73) & 3 & 0.78 (0.74, 0.81) & 2 & 0.78 (0.73, 0.81) & 2 & 0.78 (0.74, 0.81) & 2 & 0.78 (0.75, 0.82) & 1 \\ 
  3 & Age & 0.74 (0.74, 0.74) & 2 & 0.77 (0.72, 0.8) & 3 & 0.76 (0.72, 0.8) & 4 & 0.77 (0.72, 0.8) & 3 & 0.74 (0.7, 0.77) & 4 \\ 
  4 & ASTP & 0.73 (0.73, 0.73) & 4 & 0.76 (0.73, 0.8) & 4 & 0.76 (0.73, 0.81) & 3 & 0.76 (0.73, 0.8) & 4 & 0.74 (0.7, 0.78) & 3 \\ 
  5 & Albumin & 0.65 (0.65, 0.65) & 5 & 0.7 (0.66, 0.73) & 5 & 0.7 (0.66, 0.73) & 5 & 0.7 (0.66, 0.73) & 5 & 0.68 (0.64, 0.71) & 5 \\ 
  6 & Systolic BP & 0.54 (0.54, 0.54) & 6 & 0.53 (0.5, 0.57) & 6 & 0.53 (0.5, 0.57) & 6 & 0.53 (0.5, 0.57) & 6 & 0.5 (0.5, 0.57) & 6 \\ 
   \hline
\end{tabular}}
\caption{AUC estimates and $95\%$ bootstrap confidence intervals for continuous predictors in NHANES 2003-2006.}
\label{tab: auc-nhanes}
\end{table}

\begin{table}[htb!]
\centering
\resizebox{\textwidth}{!}{\begin{tabular}{|r|l|l|l|l|l|l|l|l|l|l|l|}
  \hline
 & Variables & $R_{lKuw}^2$ & Rank & $R_{lKpw}^2$ & Rank & $R_{lS}^2$ & Rank & $R_{lQ}^2$ & Rank \\ 
  \hline
1 & TAC & 0.33 (0.28, 0.39) & 1 & 0.29 (0.2, 0.37) & 1 & 0.29 (0.2, 0.37) & 1 & 0.23 (0.17, 0.29) & 2 \\ 
  2 & MVPA & 0.27 (0.23, 0.35) & 3 & 0.25 (0.18, 0.32) & 2 & 0.25 (0.18, 0.32) & 2 & 0.26 (0.19, 0.35) & 1 \\ 
  3 & Age & 0.29 (0.24, 0.36) & 2 & 0.23 (0.15, 0.3) & 3 & 0.23 (0.15, 0.3) & 3 & 0.19 (0.12, 0.24) & 4 \\ 
  4 & ASTP & 0.26 (0.22, 0.3) & 4 & 0.22 (0.16, 0.31) & 4 & 0.22 (0.16, 0.31) & 4 & 0.19 (0.13, 0.25) & 3 \\ 
  5 & Albumin & 0.11 (0.1, 0.14) & 5 & 0.13 (0.08, 0.17) & 5 & 0.13 (0.08, 0.17) & 5 & 0.1 (0.06, 0.15) & 5 \\ 
  6 & Systolic BP & 0.01 (0.01, 0.01) & 6 & 0 (0, 0.02) & 6 & 0 (0, 0.02) & 6 & 0 (0, 0.01) & 6 \\ 
   \hline
\end{tabular}}
\caption{$R^2_l$ estimates and $95\%$ confidence intervals for continuous predictors in NHANES 2003-2006.}
\label{tab: rsq-nhanes}
\end{table}

\section{Discussion}
%
We used Semiparametric Gaussian Copula to define latent $R^2_l$, a measure of variation explained for the case of observed binary outcome and observed continuous predictor. Conceptually, $R_l^2$ can be considered as a parameter of the data generating process that does not depend on the prevalence rate $p$ and have an intuitive scale-invariant interpretation. The scale-dependence was considered to be a major limitation of other previously proposed $R^2$-type measures\citep{demaris2002explained, schemper2003predictive, steyerberg2009clinical}. Under SGC, AUC and $R^2_l$ are directly related and their mutually consistent interpretation can provide a more complete description of both discrimination and dependence, especially, under highly unbalanced cases \citep{saito2015precision}. We showed that if $p = 0.5$, AUC and the latent correlation are almost linearly related. However, once $p$ gets smaller, the two measures exhibit a significant nonlinear divergence. Hence, for a fixed AUC, variation explained is getting smaller while the prevalence rate getting smaller. This is similar to examples considered in Chapter 15 of \cite{steyerberg2009clinical} that compared AUC vs likelihood-based Nagelkerke's $R^2$ for different values of the prevalence rate. Note that our proposal established an exact relationship between AUC and scale-invariant semiparametric $R^2_l$.  We also demonstrated how four rank statistics and prevalence rate can be used to estimate both AUC and the latent $R^2_l$. We proved that our weighted AUC estimators defined through Spearman and Quadrant correlations are asymptotically normal and hence, consistent under reasonable complex survey design assumptions. We additionally showed that AUC is sensitive to outliers and proposed $A_Q$, AUC calculated via Quadrant rank correlation, as a robust semiparametric version of AUC. Finally, we demonstrated how AUC can be calculated using only single participant survey weights under complex survey designs. As of interesting note, we showed that Kendall's Tau and Spearman rank correlations are linearly related in a binary-continuous case,  which is in contrast to the continuous-continuous case where they are only asymptotically equivalent and the latter is a linear projection of the former \citep{sidak1999theory}.

There are a few limitations in the proposed framework. First, the main assumption is that binary-continuous pairs are generated according to a Semiparametric Gaussian Copula. Even though SGC is a flexible framework, it is desirable to develop data-driven methods to test this assumption and be able to detect deviations from it. Another limitation, that remains to be addressed in the future work, is that we do not handle in any way the presence of ties that may occur in practice even for continuous variables.

This work can be extended in many interesting ways. First, latent continuous random variables generating observed binary outcomes could be of interest by themselves and methods development for calculating best predictors for these random variables would be welcome. Second, the SGC approach has been recently extended to include truncated variables in \citep{yoon2018sparse} and some specific cases of ordinal variables in \citep{quan2018rank}. This opens up an opportunity to extend $R^2_l$ to a wider class of mixed data types. Similarly to \cite{croux2010influence} and  \cite{nikitin1995asymptotic}, the future work should investigate and compare the asymptotic efficiency of the proposed estimators of AUC and $R^2_l$. It also would be interesting to compare $R^2_l$ to other previously proposed $R^2$-type measures. Finally, extending $R^2_l$  to the multivariate mixed data type predictors would be a natural next step.

\vspace{0.5in}
 {\bf Acknowledgment:} We would like to thank Professor Thomas Lumley for a discussion of our approach and pointing to his blog post on the use of Wilcoxon rank-sum statistics under complex designs. We would like to thank Dr. Barry Graubard and Dr. Wenliang Yao for sharing their R code with us and providing useful insights. We also would like to thank Professor Ravi Varadhan and Dr. Stas Kolenikov for early discussions of estimating AUC under complex survey designs. Finally, we would like to thank Professor Irina Gaynanova for her insightful discussions of the SGC model and its extensions.

\appendix

\appendixone
\section*{Appendix 1: Technical details and proofs}
\subsection{Derivations of the relationships between rank statistics and AUC}\label{appendix: auc-rank}

In the next discussion, we repeatedly use the fact that for continuous random variable $X$, $sgn(X)=2I(X>0)-1$ with probability one. 
\begin{equation}
\begin{split}
r_K &=  E((Y_i - Y_i') sgn(X_i - X_i')) \\ 
 &=  E(sgn(X_i - X_i') | (Y_i-Y_i') = 1) P((Y_i-Y_i') = 1) \\
  & - E(sgn(X_i- X_i') | (Y_i-Y_i') = -1) P((Y_i-Y_i') = -1) \\
  &= p(1-p)(E(sgn(X_i - X_i') | (Y_i-Y_i') = 1)-E(sgn(X_i- X_i') | (Y_i-Y_i') = -1))\\
  &=p(1-p)(2P(X_i > X_i' | (Y_i-Y_i') = 1)-2P(X_i > X_i'| (Y_i-Y_i') = -1))\\
  & = 2p(1-p)(P(X_1 > X_0) - P(X_1 < X_0)) \\
  & \implies |r_K| = 2p(1-p) |P(X_1 > X_0) - P(X_1 < X_0)|  
\end{split}
 \label{eq1}
\end{equation}

From the definition of AUC, we know that, $|P(X_1 > X_0) - P(X_1 < X_0)| = (2A-1)$. Combining this fact with Equation (\ref{eq1}), we get - 
\begin{equation}
\begin{split}
A & = \frac{1}{2} + \left|\frac{r_K}{4p(1-p)}\right|  
\end{split}
 \label{auc-kendall}
\end{equation}
\begin{equation}
\begin{split}
 W &=  P(X \leq X_1) - P(X \leq X_0) \\
 & = P(X \leq X_1|Y=0)P(Y=0) + P(X \leq X_1|Y=1)P(Y=1) - \\
 &  P(X \leq X_0|Y=0)P(Y=0) - P(X \leq X_0|Y=1)P(Y=1) \\
 & = P(X_1 > X_0)(1-p) + P(X_1' \leq X_1) p - \\
 & P(X_0' \leq X_0)(1-p) -P(X_1 < X_0) p \\
 & = P(X_1>X_0) + \frac{1}{2}p - \frac{1}{2}(1-p) - p\\
 & [X_1 \stackrel{d}{=} X_1', X_0 \stackrel{d}{=} X_0', \therefore P(X_1 \leq X_1')=P(X_0 \leq X_0')=\frac{1}{2}]\\
 & = P(X_1 > X_0) - \frac{1}{2}\\
 \implies |W| & = A - \frac{1}{2} \\
 \implies A & =  \frac{1}{2} + |W|
   \end{split}
   \label{wilcox}
\end{equation}

\begin{equation}
\begin{split}
 r_S &= 12E[F_X(X) F_Y(Y)] - 3 \\
 & = 12E[F_X(X')\{(1-p)\mathbb{I}(Y'=0) + \mathbb{I}(Y'=1)\}] - 3\\
 & = 12E[F_X(X)] - 12pE[F_X(X')\mathbb{I}(Y'=0)] - 3 \\
 & = 3 - 12p E[P(X\leq X' | X')\mathbb{I}(Y'=0)]\\
 & = 3 - 12p E[E[\mathbb{I}(X\leq X')\mathbb{I}(Y'=0)|X',Y']] \\
 & = 3 - 12p P(X \leq X', Y'=0)\\
 \implies P(X \leq X', Y'=0) & = \frac{3 - r_S}{12p}
 \end{split}\label{spear-wil1}
 \end{equation}
 
\begin{equation}
\begin{split}
 W &=  P(X \leq X_1) - P(X \leq X_0) \\
 & = P( X \leq X'|Y'=1) - P(X \leq X'|Y'=0)\\
 & = \frac{P(X \leq X', Y'=1)}{p} - \frac{P(X \leq X' , Y'=0)}{1-p}\\
 & = \frac{P(X \leq X', Y'=1) + P(X \leq X', Y'=0)}{p} - P(X \leq X', Y'=0)(\frac{1}{p} + \frac{1}{1-p}) \\
 & = \frac{1}{2p} - \frac{P(X \leq X', Y'=0)}{p(1-p)}\\
 \end{split}\label{spear-wil2}
 \end{equation}
 
 Combining equation (\ref{spear-wil1}) and (\ref{spear-wil2}), we get - 
 \begin{equation}
\begin{split}
W &= \frac{1}{2p} - \frac{\frac{3-r_S}{12p}}{p(1-p)}\\
& = \frac{r_S - (6p^2 - 6p+3)}{12p^2(1-p)}\\
\implies A & = \frac{1}{2} + \left|\frac{r_S - (6p^2 - 6p+3)}{12p^2(1-p)}\right| 
\end{split}
 \end{equation}
 
 \subsection{Derivation of bridging functions and proofs of Lemmas 1 and 2.} \label{appendix:bridge}
 We lay out the proof of Lemma \ref{lemma: sgc-bridge} below. Remember that $F_Y(y)=(1-p)\mathbb{I}(y<1)+ \mathbb{I}(y >= 1)$, so, $F_Y(Y)=(1-p)\mathbb{I}(Y<1)+ \mathbb{I}(Y >= 1)=p\mathbb{I}(U <= \Delta)+ \mathbb{I}(U > \Delta)$. Also, $F_X(X)= \Phi(V)$, $\Phi(V) \sim U(0,1)$ and $E(\Phi(V))=0.5$. Hence, 
 \begin{equation}
\begin{split}
r_S &=12E[F_Y(Y)F_X(X)] - 3 \\
& = 12 E[P(X_2 < X_1 |X_1) P(Y_3 \leq Y_1|Y_1)]- 3\\
& = 12 E[P(X_2 < X_1 |X_1) P(Y_3 \leq Y_1|Y_1)] - 3 \\
&= 12 E[E[\mathbb{I}(X_2 < X_1 , Y_3 \leq Y_1)|X_1,Y_1]] - 3\\
& = 12 P(X_2 < X_1, Y_3 \leq Y_1) - 3\\
& = 12 \gamma - 3
\end{split}
\label{type2}
\end{equation}

Using the established relationship between Spearman's rank correlation and Type 2 concordance, $\gamma$, our next step is to derive the relationship between Type 2 concordance and the latent correlation, $r$.

\begin{equation}
\begin{split}
\gamma & = P(X_2 < X_1, Y_3 \leq Y_1) \\
& = P(V_2 < V_1, U_3 <= \Delta, U_1 > \Delta] + P(V_2 < V_1, U_3 > \Delta, U_1 > \Delta) + \\
& P(V_2 < V_1, U_3 <= \Delta, U_1 <= \Delta) \\ 
& = (1-p)P(\frac{(V_2 - V_1)}{\sqrt(2)}< 0, U_1 > \Delta) + pP(\frac{(V_2 - V_1)}{\sqrt(2)}<0, U_1 > \Delta) + \\
& (1-p)P(\frac{(V_2 - V_1)}{\sqrt(2)}<0, U_1 <= \Delta) \\
& = (1-p) \Phi_2(0,-\Delta,\frac{r}{\sqrt{2}}) + p \Phi_2(0,-\Delta,\frac{r}{\sqrt{2}}) + p \Phi_2(0,\Delta,-\frac{r}{\sqrt{2}}) \\
& = \Phi_2(0,-\Delta,\frac{r}{\sqrt{2}}) + p \Phi_2(0,\Delta,-\frac{r}{\sqrt{2}}) \\ 
\end{split}
\label{type2bridge}
\end{equation}

Using Equations (\ref{type2}) and (\ref{type2bridge}), we can conclude that  
\begin{equation}
\begin{split}
r_S & = 12 [\Phi_2(0,-\Delta,\frac{r}{\sqrt{2}}) + p \Phi_2(0,\Delta,-\frac{r}{\sqrt{2}})] - 3  = G_S (r)
\end{split}
\end{equation}

We derive the bridging function for Quadrant rank correlation as follows.  

 \begin{equation}
\begin{split}
 r_Q &= E[sgn((Y-M_Y)(X-M_X))] \\
 & =  P(U > \Delta, V > 0) - P(U > \Delta, V < 0) \mathbb{I}(M_Y=0) + \\
 &  P(U < \Delta, V < 0) - P(U < \Delta, V > 0) \mathbb{I} (M_Y=1) \\
 & =  [\Phi_2(-\Delta,0,r) - \Phi_2(-\Delta,0,-r)] \mathbb{I}(M_Y=0) + \\
 &  [\Phi_2(\Delta,0,r) - \Phi_2(\Delta,0,-r)] \mathbb{I} (M_Y=1)  = G_Q(r)
   \end{split}
   \label{quadbridge}
\end{equation}

\cite{fan2017high} proved the following result in the appendix and used it to prove monotonicity and hence invertibility of $G_K(r)$. The result is stated as follows. 
\begin{lemma}
For any fixed $\Delta_1,\Delta_2$, $\Phi_2(\Delta_1,\Delta_2,r)= \int_{-\infty}^{\Delta_1} \Phi(\frac{\Delta_2 - rx}{\sqrt{1-r^2}}) \phi(x) dx$, where $\phi(x)$ is the standard normal density, and moreover, $ \Phi_2(\Delta_1,\Delta_2,r)$ is a strictly increasing function of $r$ in $(-1,1)$. Hence, inverse exists for the function. 
\end{lemma}

Following these ideas that have been used to prove the monotonicity of the bridging function $G_K(r)$, we will now prove Lemma \ref{lemma: bridge-inv}.

\begin{proof} Without the loss of generality, we assume that $M_Y=0$, and define $F_\Delta(r)=\Phi_2(-\Delta,0,r)=\Phi_2(0,-\Delta,r)$. Then, using Lemma 1, $\frac{\partial F_\Delta(r)}{\partial r} = F_\Delta'(r) > 0 $ for $r \in (-1,1)$. Also, we can write  \begin{equation}
\begin{split}
G_Q(r) & = \Phi_2(-\Delta,0,r) - \Phi_2(-\Delta,0,-r)=F_\Delta(r) - F_\Delta(-r)\\
\implies \frac{\partial G_Q(r)}{\partial r} & = F_\Delta'(r) + F_\Delta'(-r) > 0
\end{split} \label{quad-inc}
\end{equation}
Also using the fact that $\Phi_2(\Delta_1,\Delta_2,-r)=\Phi(\Delta_1) - \Phi(\Delta_1,-\Delta_2,r)$, we can derive  
\begin{equation}
\begin{split}
G_S(r) & = 12 [\Phi_2(0,-\Delta,\frac{r}{\sqrt{2}}) + p \Phi_2(0,\Delta,-\frac{r}{\sqrt{2}})] - 3\\
& = 12[(1-p)\Phi_2(0,-\Delta,\frac{r}{\sqrt{2}}) + \frac{p}{2}] - 3\\
& = 12[(1-p)F_\Delta(\frac{r}{\sqrt{2}})] + (6p-3)\\
\implies \frac{\partial G_S(r)}{\partial r} &= \frac{12(1-p)}{\sqrt{2}} F_\Delta'(\frac{r}{\sqrt{2}}) > 0. 
\end{split} \label{spear-inc}
\end{equation}
\end{proof}
This proves the statement.

\subsection{Proof of Theorem 1.}\label{appendix: asymp-qs}


We establish asymptotical results under a sequence of finite-populations $P_{N_\nu}$ converging to a super-population $P$, following the framework proposed by \cite{rubin2005two}. We assume that a finite sample of size $n_\nu$ is drawn from finite population of size $N_\nu$ according to a sampling design $p(s)$ and $N_\nu, n_\nu \rightarrow \infty$ with $limsup \frac{n_\nu}{N_\nu} < 1$ as $\nu \rightarrow \infty$. The superpopulation associated with $N_\nu$ is embedded with a probability space $(\Omega, \mathcal{F},\xi)$. Below, all the distributions and convergence refer to the joint process of first choosing a finite population from the super-population and then drawing a sample from the finite population using a probability sampling. The combination of the two levels of convergence is inline with Theorem 6.1 of \cite{rubin2005two}, where the authors considered the convergence of sample estimators defined by estimating equations. For next part of discussion, we denote the joint probability measure as $\xi p$ , where $\xi$  denotes the probability measure with respect to the model defined on finite population, and $p$ denotes the measure with respect to the design conditioning on the finite population. We lay out the assumptions $1,3,4,5$ from \cite{wang2012sample} below under which we will prove our results.

\begin{assumption}
The finite population $P_N$ consists of a sequence of i.i.d. variables $(Y_i, X_i), i=1,\cdots, N$.
\end{assumption} 

\begin{assumption}
The following conditions hold for inclusion probabilities $w(i)$  and design variance of Horvitz–Thompson estimator of the mean - (i) $K_L\leq N w(i)/n^\ast\leq K_U$
 for all $i$ , where $K_L$ and $K_U$ are positive constants, (ii) For any vector $z_i$ with finite  population moments, or equivalently, $\frac1N\sum_{i=1}^N{\Arrowvert\boldsymbol z\Arrowvert}^{2+\delta}<\infty$
where ${\Arrowvert\boldsymbol z\Arrowvert}=\sqrt{\boldsymbol z^T\boldsymbol z}$ denotes the $L_2$-norm of vector $z$, we assume
$n^\ast{\mathrm{Var}}_p{({\overline{\boldsymbol z}}_w)}\leq K_V$
for some $K_V > 0$  and $n^\ast = E_p(n)$.
\end{assumption} 

\begin{assumption}
For any $z$ with finite fourth population moment, $Var_p(\bar{z}_w)^-{\frac{1}{2}} (\bar{z}_w - \bar{z}_N)|\mathcal{F}_N$ converges to $N(0, \mathrm{I})$ with respect to $\mathcal{L}_p$ and ${\lbrack{\mathrm{Var}}_p{({\overline{\boldsymbol z}}_w)}\rbrack}^{-1}{\widehat V}_p{({\overline{\boldsymbol z}}_w)}-{\mathrm I}=O_p{(n^{\ast^{-1/2}})}$, where $\mathrm{I}$ is the identity matrix, the design variance–covariance matrix of $\bar{z}_w$, denoted by ${\lbrack{\mathrm{Var}}_p{({\overline{\boldsymbol z}}_w)}\rbrack}$, is positive definite, and
${\widehat V}_p{({\overline{\boldsymbol z}}_w)}=\frac1{N^2}\sum_{i\in A}\sum_{j\in A}\Omega_{ij}{\boldsymbol z}_i{{\boldsymbol z}_j^T}$, where $\Omega_{ij}$ means design-dependent weights associated with each pair $(i,j)$. 

Here, convergence with respect to $\mathcal{L}_p$ means ``convergence in the law of sampling design”, conditioning on the realized population, $\bar{z}_w$ denotes the Horvitz-Thompson estimate of $z$ from finite sample and $\bar{z}_N$ means the mean in the finite population. 
\end{assumption} 

\begin{assumption}
Let $D_{t,N}$ denote the set of all distinct $(i_1,i_2,\cdots,i_t)$-tuples from $P_N$. We have

\begin{equation}
\begin{split}
\underset{N\rightarrow\infty}{limsup}\frac{N^4}{n^{\ast^2}}\max_{{(i_1,i_2,i_3,i_4)}\in D_{4,N}}{\vert{\mathrm E}_p{({\mathrm I}_{i_1}-w(i_1))}{({\mathrm I}_{i_2}-w(i_2))}{({\mathrm I}_{i_3}-w(i_3))}{({\mathrm I}_{i_4}-w(i_4))}\vert}\leq M_1<\infty\\
\underset{N\rightarrow\infty}{limsup}\frac{N^3}{n^{\ast^2}}\max_{(i_1,i_2,i_3)\in D_{3,N}}{\vert{\mathrm E}_p{({\mathrm I}_{i_1}-w(i_1))^2}{({\mathrm I}_{i_2}-w(i_2))}{({\mathrm I}_{i_3}-w(i_3))}\vert}\leq M_2<\infty\\
\underset{N\rightarrow\infty}{limsup}\frac{N^2}{n^{\ast^2}}\max_{{(i_1,i_2)}\in D_{2,N}}{\vert{\mathrm E}_p{({\mathrm I}_{i_1}-w(i_1))^2}{({\mathrm I}_{i_2}-w(i_2))^2}\vert}\leq M_3<\infty
\end{split}
\end{equation}
almost surely for all populations. Here $\mathrm{I}_k$ denotes the indicator that $k$-th unit is chosen in the finite sample. 
\end{assumption} 

Now, we introduce the following notations for the Horvitz-Thompson estimators of population quantities from finite sample - 
\begin{equation}
\begin{split}
    \hat{F}_{0n}(t)&= \frac{1}{{\sum_{i:Y_i=0}\frac{1}{w(i)}}}\sum_{i:Y_i=0} \frac{1}{w(i)} I(X_i \leq t)\\
    \hat{F}_{1n}(t) & = \frac{1}{{\sum_{i:Y_i=1}\frac{1}{w(i)}}}\sum_{i:Y_i=1} \frac{1}{w(i)} I(X_i \leq t)\\
    \hat{F}_{n}(t) & = \frac{1}{{\sum_{i=1}^{n}\frac{1}{w(i)}}}\sum_{i=1}^{n} \frac{1}{w(i)} I(X_i \leq t)\\
    \hat{p}_w &=\frac{\sum_{i: =1}^{n}\frac{1}{w(i)}\mathbb{I}(Y_i=1)}{\sum_{i=1}^{n}\frac{1}{w(i)}}.
\end{split}
\label{eqn: dist-ht}
\end{equation}

We will also use the fact that $\hat{M}_Y= \mathbb{I}(\hat{p}_w > \frac{1}{2})$. We denote the true distribution functions of the random variables $X, X_0 = (X|Y=0), X_1 = (X|Y=1)$ as $F_X, F_{0X}$ and $F_{1X}$ respectively. Now, as the random variable $\mathbb{I}(Y_i=1)$ has finite fourth moment, using Assumption $A3$, similar to proof of Theorem 1 in \cite{wang2012sample}, it can be shown that, $\sqrt{n}((\hat{p}_w, \hat{F}_{0n}(t), \hat{F}_{1n}(t), \hat{F}_{n}(t))' - (p,F_{0X}(t),F_{1X}(t),F_{X}(t))')$ converges weakly to $N(0, \Sigma)$ for some covariance matrix $\Sigma$ that depends on second-order design probabilities. The convergence of $\hat{p}_w$, the Horvitz-Thompson estimator of the sample mean, is discussed thoroughly in Corollary 1.3.6.1 of \cite{fuller2011sampling}. We can treat the distribution functions as random elements in $\mathcal{D}[0,1]$, the space of all right continuous functions defined on $[0,1]$, and $\hat{p}_w$ as a constant process. Then similar to \cite{lumley2013two} and \cite{wang2012sample}, we can infer that  $\sqrt{n}((\hat{p}_w, \hat{F}_{0n}, \hat{F}_{1n}, \hat{F}_{n})'-(p,F_{0X},F_{1X},F_{X})')$ converges weakly to the Gaussian process $T=(T_{w_1},T_{w_2}, T_{w_3},T_{w_4})$ with the covariance kernel that depends on second order sampling probabilities. 
Following Example 20.12 in \cite{van2000asymptotic} and assuming $g(x)= I(x > \frac{1}{2})$, we can rewrite $\hat{r}_Q$ and $\hat{r}_S$ as follows 
\begin{equation}
    \begin{split}
\hat{r}_Q & =
\frac{1}{{\sum_{i=1}^{n}\frac{1}{w(i)}}}\sum_{i=1}^{n} \frac{1}{w(i)} sgn((Y_i-\hat{M}_Y)(X_i-\hat{M}_X))\\
& = \frac{1}{{\sum_{i:Y_i=1}\frac{1}{w(i)}}}\hat{p}_w \mathbb{I}(\hat{M}_Y =0) \sum_{i:Y_i=1} \frac{1}{w(i)} 2(\mathbb{I}(\hat{F}_X(X_i) > \frac{1}{2})-1) -\\
&\frac{1}{{\sum_{i:Y_i=0}\frac{1}{w(i)}}}(1-\hat{p}_w) \mathbb{I}(\hat{M}_Y =1) \sum_{i:Y_i=0} \frac{1}{w(i)} 2(\mathbb{I}(\hat{F}_X(X_i) > \frac{1}{2})-1)  \\
& = \frac{1}{{\sum_{i:Y_i=1}\frac{1}{w(i)}}}\hat{p}_w \mathbb{I}(\hat{p}_w \leq \frac{1}{2}) \sum_{i:Y_i=1} \frac{1}{w(i)} 2(\mathbb{I}(\hat{F}_X(X_i) > \frac{1}{2})-1) -\\
&\frac{1}{{\sum_{i:Y_i=0}\frac{1}{w(i)}}}(1-\hat{p}_w) \mathbb{I}(\hat{p}_w > \frac{1}{2}) \sum_{i:Y_i=0} \frac{1}{w(i)} 2(\mathbb{I}(\hat{F}_X(X_i) > \frac{1}{2})-1)  \\
& = \hat{p}_w \mathbb{I}(\hat{p}_w \leq \frac{1}{2}) \int g(\hat{F}_n) d\hat{F}_{1n} - (1-\hat{p}_w) \mathbb{I}(\hat{p}_w > \frac{1}{2}) \int g(\hat{F}_n) d\hat{F}_{0n} - \\
& - (\hat{p}_w \mathbb{I}(\hat{p}_w \leq \frac{1}{2}) (1-\hat{p}_w) \mathbb{I}(\hat{p}_w > \frac{1}{2}))
\end{split}\label{eqn: quad-asymp-form}
\end{equation}

\begin{equation}
\begin{split}
    \hat{r}_S &= 12  \frac{1}{{\sum_{i=1}^{n}\frac{1}{w(i)}}} \sum_{i=1}^{n} \frac{1}{w(i)} [\hat{F}_Y(Y_i)\hat{F}_X(X_i)] - 3\\
    & =12 [(1-\hat{p}_w)^2\int \hat{F}_n d\hat{F}_{0n} +  \hat{p}_w \int \hat{F}_n d\hat{F}_{1n}] - 3
\end{split}\label{eqn: spear-asymp-form}
\end{equation}

where $g(x) = I(x > \frac{1}{2})$ in Equation \ref{eqn: quad-asymp-form}. In Theorem 1 of \cite{lumley2013two}, the convergence of $\int g(\hat{F}_n) d\hat{F}_{1n} - \int g(\hat{F}_n) d\hat{F}_{0n}$ has been proved, where the function $g()$ has to be, (i) differentiable with bounded derivative and continuous on $[0,1]$, or, (ii) differentiable on $(0,1)$ and $\int_{0,1}g(y)^{2+\delta}$ is finite for some $\delta > 0$, or, (iii) an indicator function of a subinterval $(a,b)$ of $[0,1]$. Our function $g$ satisfies criteria (iii), so, proceeding similarly as the proof of Theorem 1 in \cite{lumley2013two} and applying functional delta method (\cite{kosorok2008introduction}, Theorem 12.1) to Equation \ref{eqn: quad-asymp-form}  \ref{eqn: spear-asymp-form}, we get that $\sqrt{n}(\hat{r}_Q - r_Q)$ converges to normal distribution. We should keep in mind that, the functional delta method requires Hadamard differentiability of the functional at $(p, F_{0X}, F_{1X}, F_{X}) $ and for the quadrant correlation (Equation \ref{eqn: quad-asymp-form}), the functional is not Hadamard differentiabile at $p=\frac{1}{2}$, so, we leave that special case out of the proof. 

Proving asymptotic normality of $\sqrt{n}(\hat{r}_S - r_S)$ is more straight-forward as we can apply functional delta method to Equation \ref{eqn: spear-asymp-form} and it will follow immediately. 

\section*{Appendix: Additional Figures}\label{appendix: fig}

\begin{figure}[H]
\centering
\includegraphics[width=130mm, height=130mm]{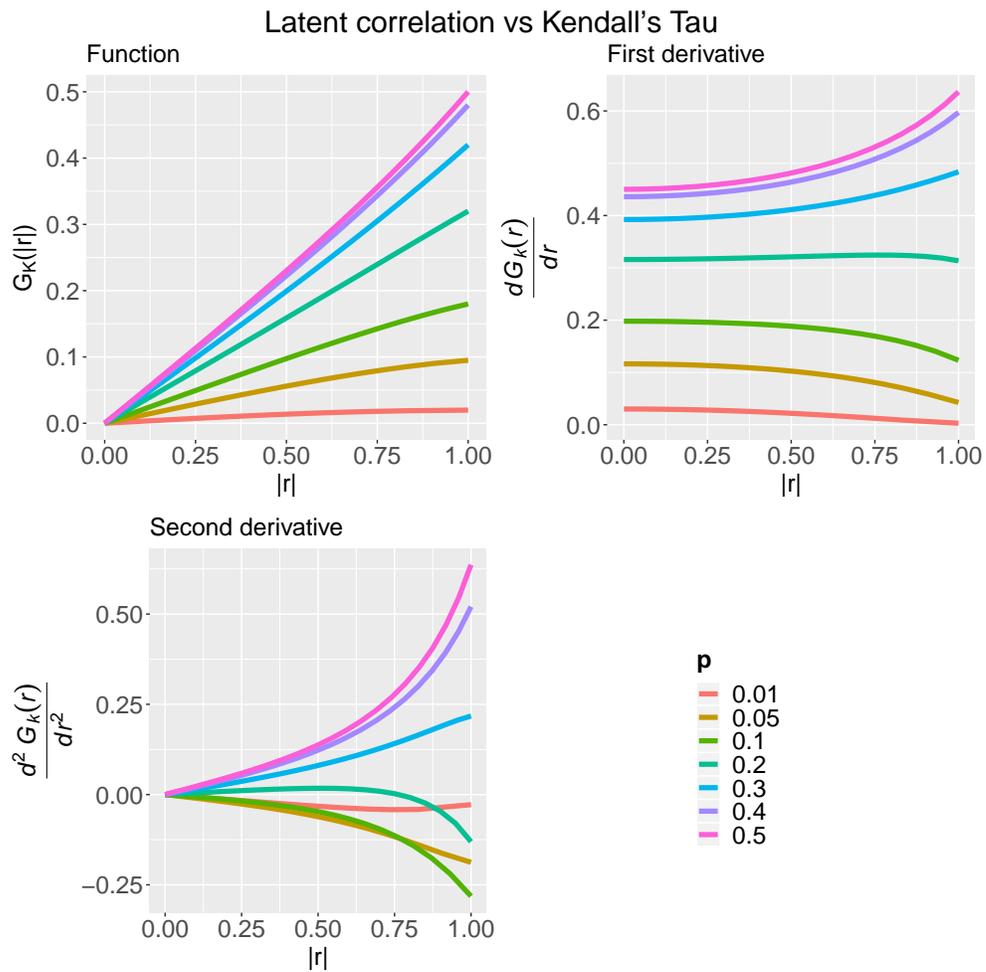}
\caption{The relationships between the absolute value of latent correlation and Kendall's Tau, varying on $p$.}
\label{fig:rkvsr}
\end{figure} 

\begin{figure}[H]
\centering
\includegraphics[width=160mm, height=100mm]{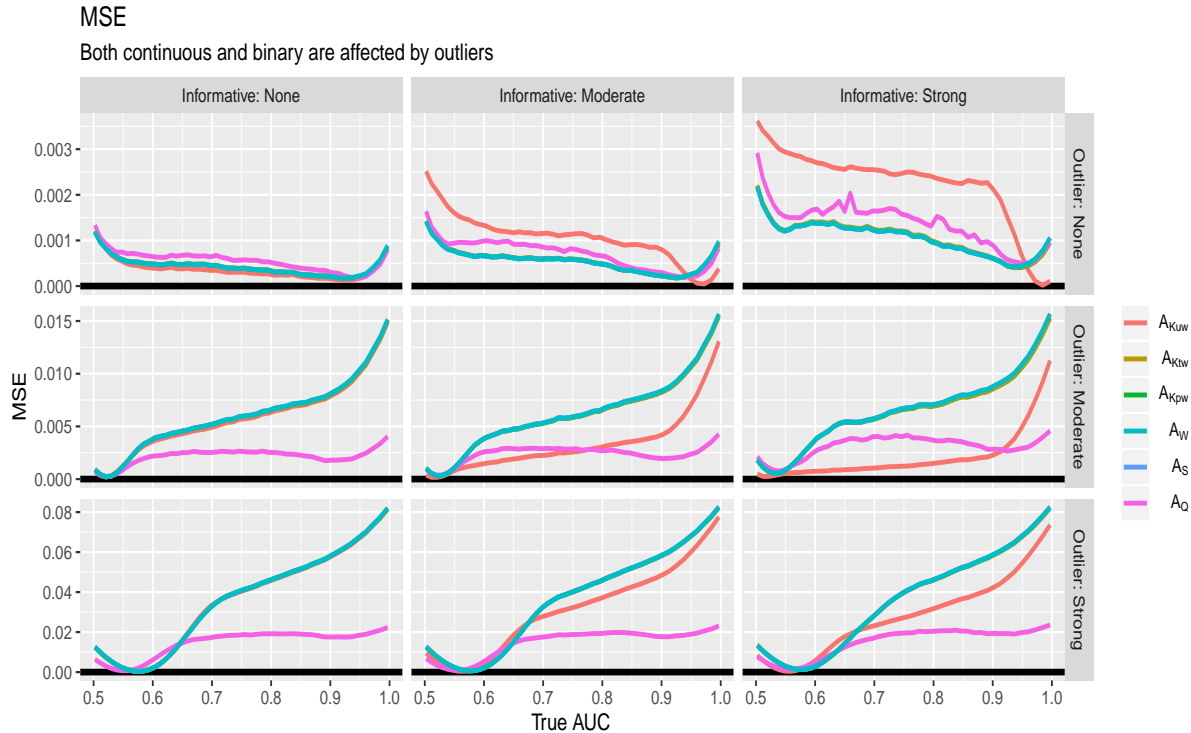}
\caption{Simulation results: MSE of the estimators under different scenario}\label{fig:mse}
\end{figure}

\bibliographystyle{biometrika}
\bibliography{ref}




.



\end{document}